# On demand single photon source using nanoscale metal-insulator-semiconductor capacitor

Binhui Hu[1], C. H. Yang[1], and M. J. Yang[2]

[1]Department of Electrical and Computer Engineering, University of Maryland, College Park, MD 20742

[2]Naval Research Laboratory, Washington DC 20375

We propose an on-demand single photon source for quantum cryptography using a metal-insulator-semiconductor quantum dot capacitor structure. The main component in the semiconductor is a p-doped quantum well, and the cylindrical gate under consideration is only nanometers in diameter. As in conventional metal-insulator-semiconductor capacitors, our system can also be biased into the inversion regime. However, due to the small gate area, at the onset of inversion there are only a few electrons residing in a quantum dot. In addition, because of the strong size quantization and large Coulomb energy, the number of electrons can be precisely controlled by the gate voltage. After holding just one electron in the inversion layer, the capacitor is quickly biased back to the flat band condition, and the subsequent radiative recombination across the bandgap results in single photon emission. We present numerical simulation results of a semiconductor heterojunction and discuss the merits of this single photon source.

73.40.Qv, 73.23.Hk, 73.63.Kv, 03.67.Dd

# 1. Introduction

Information security of conventional cryptography systems relies on the computational difficulty in factoring large numbers. In contrast, quantum cryptography, governed by the laws of quantum mechanics, is unconditionally secure,[1] and it should bring fundamental impact to future telecommunications. To implement BB84[2] quantum key distribution, it is desirable to have a single-photon source capable of producing an "on demand" single photon of known polarization in response to an external trigger.

Currently, commercial lasers with strong attenuation are used for quantum cryptography experiments. However, because the photon flux follows a Poisson distribution, there is a certain probability of having two or more photons per single pulse, giving the possibilities for eavesdropping. In order to lower the risk of emitting two or more photons per pulse, the current practice is to dramatically reduce the light intensity. Nonetheless, this approach leaves many inactive (no photon) pulses. Alternative approaches have recently attracted much interest, including parametric down-conversion,[3] a semiconductor self-assembled quantum dot in an optical microcavity,[4,5] a semiconductor self-assembled quantum dot in an LED structure,[67] trapped ions in a cavity, substitutional nitrogen vacancies in diamond,[8] single molecule luminescence,[9] a single electron-hole turnstile,[10] and using surface acoustic waves to inject single electrons.[11] The common goal is to find an on-demand single photon source with high efficiency. Recent experimental advances have already demonstrated single photon emission[3-10] and high speed operation up to 100MHz.[7]

We present in this paper a different approach using direct-bandgap semiconductors in a nanometer-scale metal-insulator-semiconductor capacitor (n-MISc) structure. The n-MISc is first biased to define a single electron quantum dot in p-type semiconductor quantum well, and subsequently, it is brought to flat band condition to trigger a radiative recombination with a hole, producing a single photon. This process repeats itself at a high rate and can be applied in quantum key distribution. In the following, we will describe in detail the structure and the operating principle of our single photon source. A numerical simulation with realistic physical parameters will also be discussed.

## 2. Device structure

We first discuss how a single electron is generated. A sketch of the proposed n-MISc is shown in Fig. 1 (a). The metallic gate is cylindrical and nanometer-scale in diameter. Below the metallic gate is the semiconductor heterostructure, including a gate dielectric, a semiconductor quantum well, an intrinsic buffer layer by the same material as used for the gate dielectric, and a heavily-doped semiconductor substrate. For photon emission, any direct bandgap semiconductor in principle can be used for the quantum well. In particular, several heterojunction systems in MIS capacitor structures have been demonstrated that they can be switched between accumulation and inversion. We have simulated $In_{0.53}Ga_{0.47}As/In_{0.52}Al_{0.48}As$,[12] and $GaAs/Al_{0.3}Ga_{0.7}As$ heterojunction systems. The results are qualitatively the same; so only the simulation from the $In_{0.53}Ga_{0.47}As$ system is discussed here. Note that for the GaAs/AlAs system, the gate dielectric can be replaced by epitaxial $Ga_2O_3$, which has recently been used in complimentary MIS field-effect transistors.[13] The interface between $Ga_2O_3$ and GaAs has a low density of interface states (~$3.5\times10^{10}cm^{-2}$),[14] while for $GaAs/Al_xGa_{1-x}As$ heterojunctions, the interface density of states is in the $10^9cm^{-2}$ range. However, $Ga_2O_3$ can sustain a higher electrical field.

## 3. Numerical simulation

In a conventional MIS capacitor with a p-type semiconductor, a gate bias can change the system from an accumulation of holes to an inversion layer of electrons.[15] With a nano-scale gate area, the inversion electrons will reside in a quantum dot. To understand the potential profile that confines the inversion electrons better, we have carried out a numerical simulation. Several assumptions and boundary conditions are detailed below. The metallic gate (see Fig. 1(a)), 150nm tall and 70nm in diameter, is biased at the gate voltage ($V_{gate}$) relative to the Fermi level in the conductive quantum well and the conductive substrate. The gate dielectric (undoped $In_{0.52}Al_{0.48}As$), the quantum well (p-doped $In_{0.53}Ga_{0.47}As$), the buffer (undoped $In_{0.52}Al_{0.48}As$), and the substrate (doped InP) have thicknesses of 50nm, 10nm, 300nm, and 1000nm respectively. Their static dielectric constant is 14. The semiconductor mesa has an area of 1μm by 1μm. All of the exposed semiconductor surfaces are assumed to have no net charges. The p-type $In_{0.53}Ga_{0.47}As$ quantum well has a bandgap of 0.75eV and is p-doped to $1\times10^{11}$ holes/$cm^2$, and its electron and heavy hole masses are $0.045m_o$ and $0.38m_o$, respectively. In the $In_{0.53}Ga_{0.47}As$ quantum well, at low temperatures, the degenerate holes have a Fermi energy of 0.63meV. For a particular $V_{gate}$, we solve[16] the Poisson equation:

$$\nabla^2 V(x,y,z) = -\frac{\rho(x,y,z)}{\varepsilon(x,y,z)},$$ where $\varepsilon(x,y,z)$ is the local dielectric constant and $\rho(x,y,z)$ is the net charge, including the free holes, free electrons, and the ionized acceptors. The solution is the potential $V(x,y,z)$ as a function of $V_{gate}$.

We focus here on the band bending in the quantum well region below the gate. At zero gate voltage, the n-MISc is in the accumulation regime. As $V_{gate}$ increases, the conduction band minimum in the quantum well is lowered and moves toward the Fermi level. We find that when $V_{gate} \sim 2.56$V, the bottom of the conduction band minimum, $E_c$ is aligned to the Fermi level. As shown in Fig. 1(b), the lateral potential profile defining the inversion layer is cylindrically symmetric and approximately parabolic.[17] Using a two-dimensional parabolic confinement potential, shown in Fig. 1(c), the eigen energy of electrons is calculated to be $\hbar\omega_0 = 12.5$meV. The parabolic confinement has an energy ladder[18] of $\hbar\omega_0(n + |\ell| + 1)$, where n = 0, 1, 2,…, and $\ell$ = 0, ±1, ±2,… When $V_{gate}$ is increased to ~ 2.63V, the Fermi level is aligned to the ground state of the quantum dot, defining the onset of the population of the first electron ($V_{gate} = V_{onset}$). Our simulation indicates that it takes a change in $V_{gate}$ of ~65mV to move the Fermi level in the induced quantum dot by 12.5meV, a factor of 5 in gating effectiveness. After the first electron occupies the ground state, the Coulomb repulsion prevents the second electron from entering the dot with the same energy, a phenomenon known as the Coulomb blockade. As observed in similar quantum dot systems, ours also has a single electron charging energy of a few meV.[17,19,20] In other words, near the onset of inversion, the number of electrons induced in the inversion layer can be well controlled by the gate bias. Such a precise control over the number of electrons has been demonstrated in both vertical[17,19] and lateral single electron transistor[20] structures.

## 4. Operating principle

At room temperature, as the gate voltage is changing, the system moves toward equilibrium by thermal generation-recombination processes. In contrast, at low temperatures, the formation of inversion electrons in this n-MISc is made possible by tunneling. Note that the surrounding p-type quantum well is filled by two-dimensional holes, whose Fermi level is kept at zero. Single electrons tunnel back and forth between the discrete states in the quantum dot and those at the Fermi level in the surrounding quantum well, when their energy levels are aligned. As in the case of Esaki diodes, this tunneling rate is of the order of 1GHz. In fact, tunneling diodes are widely used as microwave detectors for their high speed performance.

Single photons are generated by operating the n-MISc in the following sequence. Initially, the n-MISc system is reset to the flat band condition ($V_{gate} = V_{FB} = 0$ V in our assumption). The electron concentration is zero everywhere in the quantum well under equilibrium at low temperature. Then, a positive gate voltage brings the lowest quantum dot state below the Fermi level by an amount smaller than the single electron charging energy, e.g., $V_{gate} = V_{onset} + 5$mV. As a result, a single electron is induced in the quantum dot through tunneling. As indicated in Fig. 2, the duration of this process, $T_1$, must be longer than the tunneling time, $\tau_t$ (~1ns). After populating the quantum dot by one electron, the n-MISc is abruptly, within a short time $T_2$, returned to the flat band condition ($V_{gate} = V_{FB}$) again. Finally, the system is kept at flat band condition for time $T_3$. The induced single electron will stay at the conduction band before it finally recombines with a hole in the p-type quantum well and emits a single photon. The typical radiative recombination lifetime ($\tau_{rad}$) in GaAs quantum nanostructures is less than 1ns,[21] and our design calls for $T3 >> \tau_{rad}$.

## 5. Discussion

We can estimate several key characteristics. Within each gate pulse cycle, the system evolves from reset to inversion and back to reset. The repetition rate is therefore $1/(T_1 + T_2 + T_3)$. Using a conservative estimate, where $T1$ = 10ns (assumed ~10 $\tau_t$), $T2$ = 0.1ns (limited by power supply), and $T3$ = 10ns (assumed ~10 $\tau_{rad}$), the repetition rate is 50MHz. A higher rate is possible if tunneling time is reduced or the recombination rate becomes faster. Radiative recombination is a stochastic process, and it competes with non-radiative recombination. For electrons in p-type GaAs, the non-radiative recombination lifetime, $\tau_{non}$, can be prolonged by eliminating defects near mid-bandgap. In fact, recently, there has been an experimental demonstration of the total absence of non-radiative recombination[22] in modulation doped GaAs heterojunctions. In addition, we anticipate that Auger recombination is also negligible, because of the relatively low free carrier concentration. Therefore, the single electron to single photon conversion ratio $(1/\tau_{rad})/[1/\tau_{rad} + 1/\tau_{non}] = \tau_{non}/(\tau_{non} + \tau_{rad})$ is approaching 100%. The emission energy can be tuned for fiber transmission (1300 and 1550 nm pass bands) by adjusting the $In_xGa_{1-x}As$ quantum well thickness.

The spectral purity of our devices may not be as good as that of self-assembled InAs quantum dots, because the radiative recombination is a stochastic process and photon emission might occur while the system is still settling down to the reset condition. However, a high spectral purity is not required for BB84[2] quantum key distribution.

In BB84[2] quantum key distribution, the quantum bits are encoded using the polarization of photons. The polarization of the photons using the n-MISc can be manipulated by controlling the spin of the single electron and the background holes through either static (Zeeman effect) or pulsed (Rabi rotation) magnetic fields. Figure 3 depicts the Zeeman splitting of the lowest electron and hole states as a function of magnetic field. With $In_{0.53}Ga_{0.47}As$ electron g factor = -3, the Zeeman splitting is $g \cdot \mu_B \cdot B_z \cdot s_z$ = 0.87meV at 5 tesla. With the magnetic field direction normal to the plane of the two-dimensional $In_{0.53}Ga_{0.47}As$ quantum well, the (electron) ground state luminescence (i.e., electron spin +1/2, heavy hole spin +3/2) is left circularly polarized[23] (LCP, $\sigma^-$). Quantum selection rule forbids the electron spin with +1/2 to recombine with heavy hole -3/2 through dipole coupling. Furthermore, an ac magnetic field in a direction parallel to the quantum well can rotate the electron spin. For example, a $\pi$ pulse would flip the electron spin to -1/2, which has to recombine with a spin -3/2 heavy hole and emit a right circularly polarized (RCP, $\sigma^+$) photon. By controlling the strength and duration of the pulsed magnetic field, the polarization of the photons can therefore be manipulated. The polarization results from the Zeeman effect which is a fundamental property of semiconductors. This is an advantage, because the polarization of single photon sources using self-assembled InAs quantum dots relies on the incidental asymmetry of the shape, size, orientation, and uncertain alloy composition.

In summary, we have proposed a novel scheme to generate single photons using nanometer-scale MIS capacitors. We have carried out numerical simulations to demonstrate how a single electron can be generated in the inversion layer, and then recombine with a hole in the p-type quantum well. Using realistic physical parameters, we estimate that the repetition rate ought to be as high as 50MHz. The nanometer-scale gates can be readily fabricated by electron-beam lithography, metal evaporation, and lift-off. The ease of fabrication promises a massive array for a versatile operation.

Acknowlegement: This work was supported in part by Laboratory for Physical Sciences, National Security Agency, and Office of Naval Research. The authors are grateful to discussions with Dr. Dan Gammon at Naval Research Laboratory.

**Figure Captions**

Fig. 1: (a) A schematic cross-section of an n-MISc capacitor. Various components of the capacitor structure are illustrated. (b) The conduction band minimum in the quantum well below the gate is plotted as a function of the two lateral directions, $x$ and $y$. For clarity, the density of the mesh points is reduced by a factor of 10. The zero in potential is aligned to the Fermi level of holes in the quantum well. (c) The potential profile near the bottom of the two-dimensional parabolic confinement: the squares results from simulation and the best fit (line) is used for calculating the eigen energy of a two-dimensional harmonic oscillator. Orbital confinement results in a sequence of quantized states, illustrated as short bars at 12.5meV and 25 meV, and each state (bar) has a degeneracy of two.

Fig. 2: Schematics that illustrate the changes in band bending and the gate bias, at different phases of the operation. (a) Reset to flat-band condition. (b) Under positive gate voltage, a single electron tunnels into the lowest single electron state. (d) The gate bias returns to the flat band condition, and the single electron recombines with a hole in the valence band. (d) The gate voltage, $V_{gate}$, cycles through periods $T_1$, $T_2$, and $T_3$.

Fig. 3: Zeeman effect of the single electron and the two-dimensional heavy holes. The Zeeman splitting for electrons ($g_e^* \mu_B B_o S_z$, where $g_e^* < 0$, and $|S_z| = 1/2$) and holes ($g_h^* \mu_B B_o S_z$, where $g_h^* > 0$, and $|S_z| = 3/2$) for two-dimensional quantum wells allows for transitions $\sigma^-$ (left circularly polarized, electron $|+1/2\rangle$ to hole $|+3/2\rangle$) and $\sigma^+$ (right circularly polarized, electron $|-1/2\rangle$ to hole $|-3/2\rangle$).

Figure 1

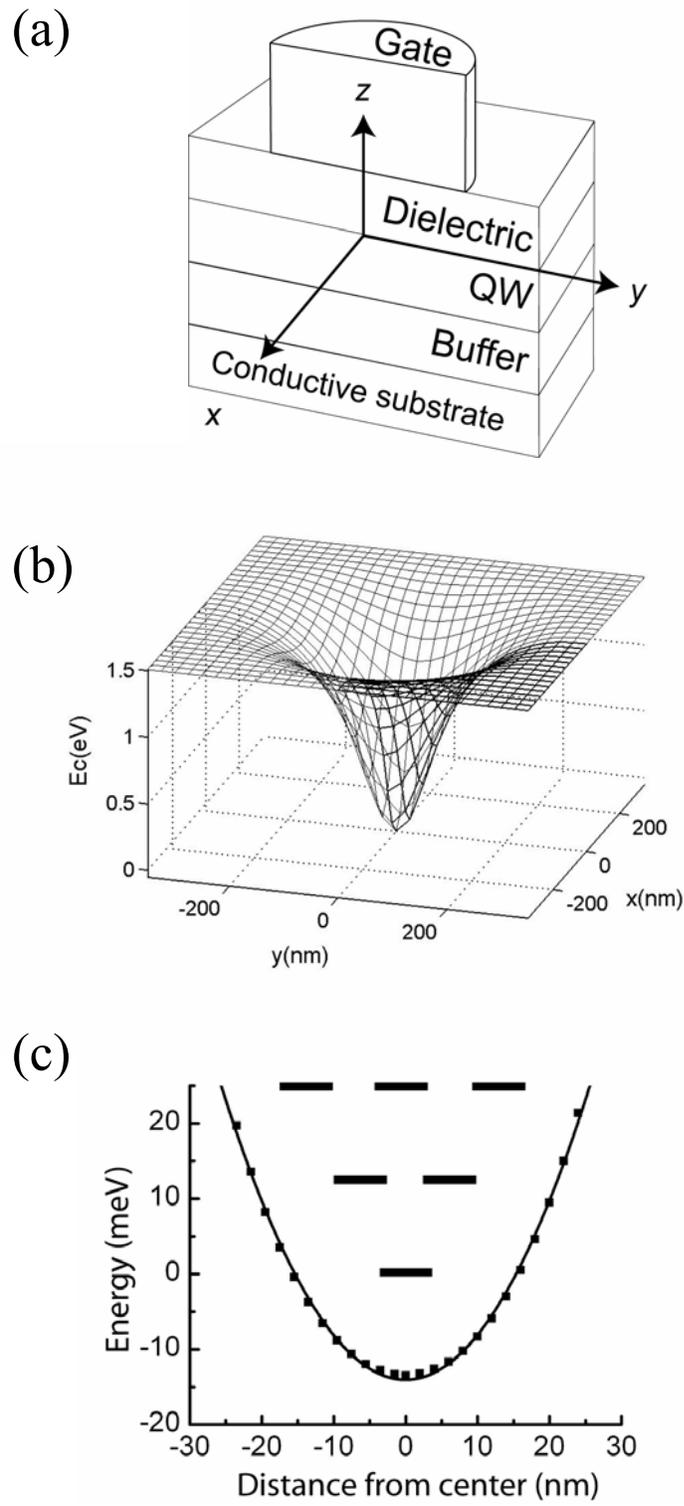

**Figure 2**

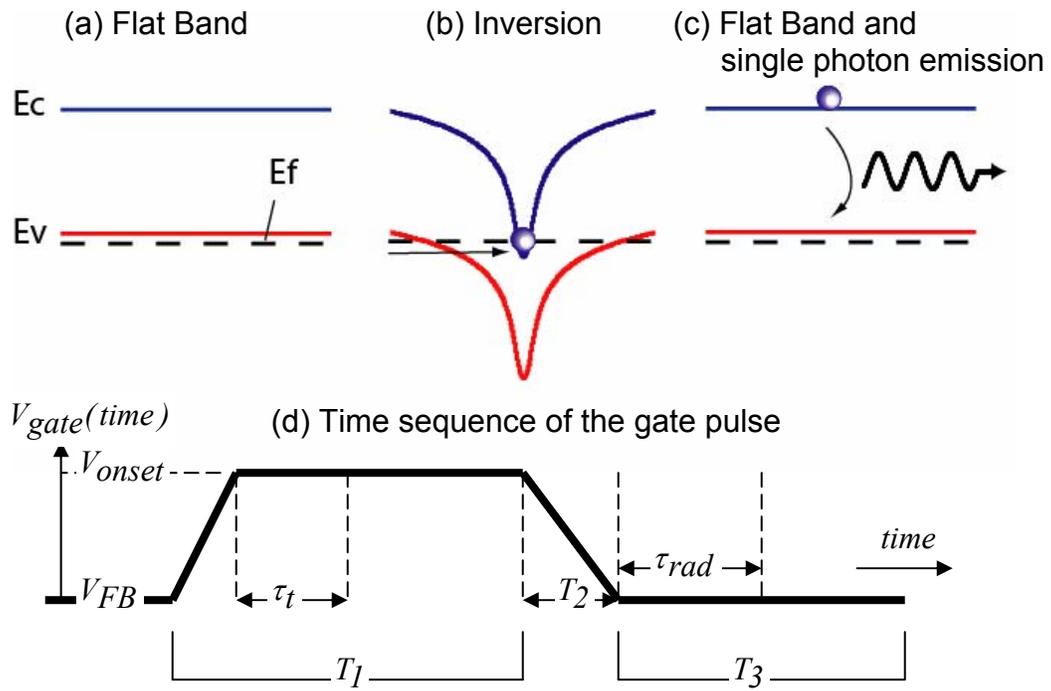

**Figure 3**

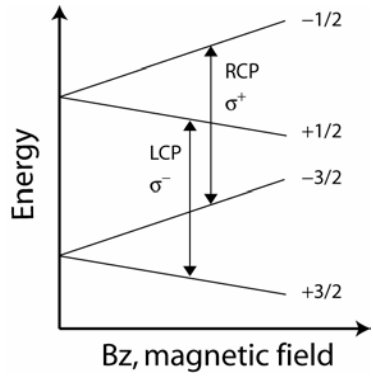